\documentclass[aps,prb,reprint,groupedaddress,floatfix,superscriptaddress,longbibliography,nofootinbib]{revtex4-2}
\usepackage[colorlinks = true,
            linkcolor = red,
            urlcolor  = red,
            citecolor = blue,
            anchorcolor = blue]{hyperref}
            
\usepackage{epsfig}
\usepackage{amsmath,amssymb}
\usepackage{graphicx}
\usepackage[dvipsnames,usenames]{xcolor}
\usepackage[normalem]{ulem}
\usepackage{todonotes}
\usepackage{soul} 
\usepackage{orcidlink}
\usepackage{glossaries}
\newacronym{DQMC}{DQMC}{determinant quantum Monte Carlo}
\newacronym{eph}{$e$-ph}{electron-phonon}
\newacronym{PLD}{PLD}{pulsed laser deposition}
\newacronym{MBE}{MBE}{molecular-beam epitaxy}
\newacronym{DCA}{DCA}{dynamical cluster approximation}
\newacronym{BEC}{BEC}{Bose-Einstein condensate}
\newacronym{QMC}{QMC}{quantum Monte Carlo}
\newacronym{2D}{2D}{two-dimensional}

\newcommand{\Tc}{T_{\mathrm{c}}}
\newcommand{\Ds}{D_{\mathrm{s}}}
\newcommand{\rhos}{\rho_{\mathrm{s}}}
\newcommand{\mrm}[1]{\mathrm{#1}}
\newcommand{\tperp}{t_{\perp}}

\begin{document}



\title{Optimizing the Critical Temperature and Superfluid Density of a Metal-Superconductor Bilayer}

\author{Yutan~Zhang\orcidlink{0009-0005-9759-1613}}
\affiliation{Department of Physics and Astronomy, University of California, Davis, California 95616, USA}

\author{Philip~M.~Dee\orcidlink{0000-0002-4249-9036}}
\affiliation{Computational Sciences and Engineering Division, Oak Ridge National 
Laboratory, Oak Ridge, Tennessee 37831, USA}
\affiliation{Department of Physics and Astronomy, The University of Tennessee, Knoxville, TN 37996, USA}
\affiliation{Institute for Advanced Materials and Manufacturing, The University of Tennessee, Knoxville, TN 37996, USA\looseness=-1} 

\author{Benjamin~Cohen-Stead\orcidlink{0000-0002-7915-6280}}
\affiliation{Department of Physics and Astronomy, The University of Tennessee, Knoxville, TN 37996, USA}
\affiliation{Institute for Advanced Materials and Manufacturing, The University of Tennessee, Knoxville, TN 37996, USA\looseness=-1} 

\author{Thomas~A.~Maier\orcidlink{0000-0002-1424-9996}}
\affiliation{Computational Sciences and Engineering Division, Oak Ridge National 
Laboratory, Oak Ridge, Tennessee 37831, USA}

\author{Steven~Johnston\orcidlink{0000-0002-2343-0113}}
\affiliation{Department of Physics and Astronomy, The University of Tennessee, Knoxville, TN 37996, USA}
\affiliation{Institute for Advanced Materials and Manufacturing, The University of Tennessee, Knoxville, TN 37996, USA\looseness=-1} 

\author{Richard~Scalettar\orcidlink{0000-0002-0521-3692}}
\affiliation{Department of Physics and Astronomy, University of California, Davis, California 95616, USA}

\begin{abstract}
A promising path to realizing higher superconducting transition temperatures $\Tc$ is the strategic engineering of artificial heterostructures. 
For example, quantum materials 
could, in principle, be coupled with other materials 
to produce a more robust  superconducting state.
In this work, we add numerical support to the hypothesis that a strongly interacting
 superconductor weakened by phase fluctuations can boost its $\Tc$ by hybridizing the system with a metal. 
Using determinant quantum Monte Carlo (DQMC), we simulate a two-dimensional bilayer composed of an attractive 
Hubbard model and a metallic layer in two regimes of the interaction strength $-|U|$. In the strongly
 interacting regime, we find that increasing the interlayer hybridization $\tperp$ results in a nonmonotonic 
enhancement of $\Tc$, with an optimal value comparable to the maximum $\Tc$ observed in the single-layer
 attractive Hubbard model, confirming trends inferred from other approaches. In the intermediate coupling regime, 
when $-|U|$ is close to the value associated with the maximum $\Tc$ of the single-layer model, increasing 
$\tperp$ tends to decrease $\Tc$, implying that the correlated layer was already optimally tuned. Importantly, 
we demonstrate that the mechanism behind these trends is related to enhancement in the superfluid stiffness, 
as was initially proposed by Kivelson [Physica B: Condensed Matter {\bf 318}, 61 (2002)]. 
\end{abstract} 

\maketitle

\section{Introduction}  \label{sec: intro}

Advances in heterostructure and thin film growth techniques \cite{Tsymbal2012multifunctional, Shepelin2023practicalguide, Ha2024thinfilm} have enabled the precise coupling of distinct quantum materials within artificial heterostructures and ultra-thin films, opening new avenues for designing and exploring interfacial functionalities with unique properties. These developments have motivated attempts to engineer superconductivity in composite systems to enhance the transition temperature $T_\mathrm{c}$ and realize novel superconducting states rare in bulk superconductors. A well-known example of this is FeSe monolayers grown on oxide substrates like SrTiO$_3$~\cite{Wang2012interface, Bozovic2014new, Huang2017monolayer}, where $\Tc$ is enhanced significantly over bulk FeSe. Similarly, a third of a monolayer of Sn grown on heavily boron-doped Si(111) substrates superconducts with a $\Tc$ higher than that of Sn thin films~\cite{Wu2020superconductivity} and shows evidence of unconventional chiral $d$-wave pairing~\cite{Ming2023evidence}. More broadly, interfacing conventional $s$-wave superconductors with nontrivial topologically materials is also being pursued as a pathway toward topological superconductivity~\cite{Fu2008superconducting, Lutchyn2010Majorana, Hemian2024interface}. 

Interfacing superconductors with energetically large pairing interactions with other materials may also provide a means to enhance superconductivity. These systems often have sizable phase fluctuations, which drive the superconducting $\Tc$ far below the mean-field prediction ($\Tc^{\mathrm{MF}}$). Kivelson~\cite{Kivelson2002} has proposed that coupling such a phase fluctuation-challenged superconductor to a metal could enhance the effective superfluid stiffness $\Ds$ and that $\Tc$ could thus be driven closer to its mean-field value. Different models have since been proposed to test this proposal, using various theoretical and computational techniques~\cite{berg2008route, wachtel2012superfluid, zujev2014pairing, dee2022enhancing, bilayerattrhub}. However, combined, these methods have yet to paint a complete picture as to when and under what conditions this strategy can be used to enhance superconductivity.

\begin {figure}[t]
\includegraphics[width=0.48\textwidth]{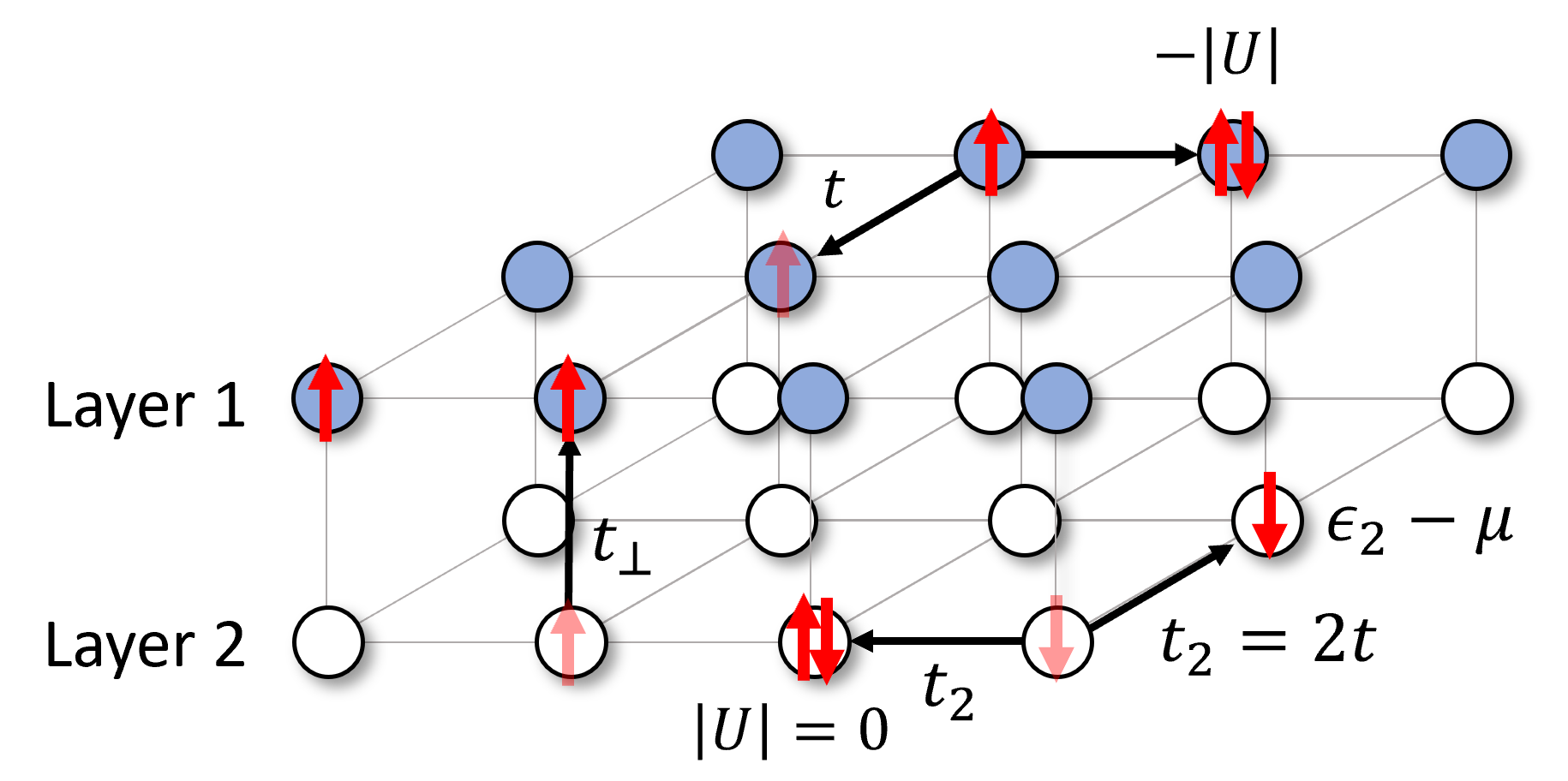}
\caption{A depiction of the bilayer model studied in this work. The top layer (layer $1$) is the attractive Hubbard layer, with a negative interaction $U$. The bottom layer (layer 2) is a metal layer with no interaction ($U=0$). The hopping of layer $1$, $t_1=t$ is set as the unit of energy throughout the work. The hopping of layer $2$, $t_2=2t$. An on-site energy $\epsilon_2=0.2t$ is set on layer $2$, to enable direct comparison with existing work. The inter-layer hopping $t_\perp$ is variable. We investigate how changing $t_\perp$ affects $\Tc$ and the physics of the system. }
\label{fig:lattice}
\end{figure}

Here we present a \gls*{DQMC} study of a specific \gls*{2D} bilayer variant of one of the models mentioned above, where a noninteracting (metallic) plane is coupled to an attractive $-|U|$ Hubbard plane through an inter-layer hopping $t_\perp$, as shown in Fig.~\ref{fig:lattice}. This model has recently been studied by some of the current authors in the strong coupling regime $|U|\sim W$, where $W$ is the bandwidth of the correlated layer, using the \gls*{DCA}~\cite{dee2022enhancing}. Their results suggest that $\Tc$ has a nonmonotonic dependence on the inter-layer hopping $t_\perp$ and is enhanced beyond the single-layer system. This behavior was explained by the competing effects of the increased pair-field susceptibility and reduced effective interaction in the correlated layer as $t_\perp$ increases. However, while the \gls*{DCA} incorporated long-range physics into a finite cluster~\cite{hettler2000dynamical,maier2005quantum}, it faced the challenge of long autocorrelation time and strong system size dependence for systems at small values of $t_\perp$. These issues made it difficult to obtain accurate $\Tc$ estimates in this limit and determine whether $\Tc$ was enhanced over that of the isolated layer. 

We address this issue in this article by studying the model~\cite{dee2022enhancing} with \gls*{DQMC}~\cite{smoqy, smoqy_code}. In particular, we obtain $\Tc$ for both strong ($U = -10t$) and intermediate ($U=-5t$) coupling regimes ($t$ is the nearest-neighbor hopping in the correlated layer). 
We do this by calculating the superfluid stiffness $\Ds$ as a function of temperature and finding its intersection with the line of $2T/\pi$~\cite{SWZ}, as we expect a Berezinskii-Kosterlitz-Thouless (BKT) type transition. This analysis is based on the fact that superfluid density is known to have a universal value $\Ds(T = \Tc) = 2\Tc/\pi$ for systems belonging
to the 2D XY universality class. Previous studies~\cite{attrHub_BCS-BEC} have investigated this approach for the single-layer attractive Hubbard model, reporting results that are only weakly dependent on system size. This makes the method well suited for addressing concerns regarding strong system size dependence in prior \gls*{DCA} work~\cite{dee2022enhancing} and allows us to definitively conclude that $\Tc$ is a nonmonotonic function of $t_\perp$ in the strong coupling regime of the bilayer model. 
We also extend the results of Ref.~\cite{dee2022enhancing} in several key ways that shed further light on the mechanisms behind the observed $\Tc$ enhancements. First, we consider the intermediate coupling regime in the correlated layer ($U =-5t$), where the attractive single-band Hubbard model has the highest $\Tc$~\cite{attrHub_BCS-BEC}. The correlated layer is less affected by phase fluctuations in this regime, and we find that the $\Tc$ of the composite bilayer system decreases monotonically with $t_\perp$. Second, we study the layer-resolved pairing structure factors, which provide insights into the nature of pairing in each layer. Finally, we find that the pair-field susceptibility and superfluid density of the $U=-10t$ and $U=-5t$ bilayer systems are enhanced compared to their optimal monolayer counterparts. However, despite those enhancements, we found the maximum $T_c$ for the bilayer to be only on par with the optimal monolayer. Whether it can be higher calls for a careful search through the parameter space and would be an exciting direction for further studies.

\section{Model \& Methods}  \label{sec: model}
\subsection{Model}
We study a bilayer square lattice Hubbard Hamiltonian, as sketched in Fig.~\ref{fig:lattice}. The model couples a strongly correlated layer ($l=1$) with a large negative-$U$ Hubbard interaction to a noninteracting metallic layer ($l=2$). The correlated layer provides strong local ($s$-wave) pairing when $|U|/t\gg1$ but has very little superfluid stiffness on its own in this regime. Conversely, the metallic layer has no intrinsic pairing interaction but a high superfluid stiffness set by its density of states. 

The model's Hamiltonian is 
\begin{align}\label{eq:hamiltonian}
 \hat{H} =&-\sum_{\langle i j\rangle, l, \sigma} t_l^{\phantom\dagger} 
 \hat{c}_{i,l,\sigma}^{\dagger} \hat{c}^{\phantom\dagger}_{j,l,\sigma}
 -t_{\perp} \sum_{i, \sigma}\left(\hat{c}_{i,1,\sigma}^{\dagger} \hat{c}^{\phantom\dagger}_{i,2,\sigma}+\text {H.c.}\right)
 \notag \\
 & +\sum_{i, l, \sigma}\left(\epsilon_l -\mu\right) \hat{n}_{i,l,\sigma}-|U| \sum_i \hat{n}_{i,1,\uparrow} \hat{n}_{i,1,\downarrow} , 
\end{align}
where $\hat{c}_{i,l,\sigma}^{\dagger}$ $(\hat{c}_{i,l,\sigma}^{\phantom\dagger})$ creates (annihilates) a spin $\sigma~(= \uparrow, \downarrow)$ electron on the $i$th site of layer $l~(= 1,2)$; $t_l$ is the in-plane nearest-neighbor hopping integral for layer $l$;  $t_\perp$ is the inter-layer hopping integral; $\mu$ is the chemical potential; and $\epsilon_l$ is an additional on-site energy term in each layer. Throughout, we set $t_1 \equiv t$ as our energy scale and fix $t_2 = 2t$, $\epsilon_1 = 0$, and $\epsilon_2 = 0.2t$ and set $\mu$ such that $n_1 \equiv \sum_\sigma \langle \hat{n}_{i,1,\sigma} \rangle=0.75$. The resulting value of $n_2=\sum_\sigma \langle \hat{n}_{i,2,\sigma} \rangle$ depends on the model parameters and temperature but typically varies between $0.9$ ($t_\perp=0$) and $0.95$ ($t_\perp=3t$) at $\beta = 10/t$. 
Our choice of parameters serves several purposes. 
First, it facilitates direct comparisons with previous \gls*{DCA} work~\cite{dee2022enhancing}. It also avoids complications of a perfectly nested Fermi surface in the metallic layer and the suppression of superconductivity in the correlated layer at half-filling. Finally, it situates the van Hove singularity of the metallic layer slightly above the chemical potential $\mu$. 

The monolayer attractive Hubbard model develops a superconducting dome as a function of doping away from half-filling, with $|U|=4t-6t$ giving the largest $\Tc$~\cite{attrHub_BCS-BEC}. The suppression of $\Tc$ for large $|U|/t$ can be intuitively understood by mapping the attractive Hubbard model to the repulsive Hubbard model by $\hat{c}_{i,l,\downarrow}^{\dagger}\to (-1)^{(i_x+i_y)}\hat{c}_{i,l,\downarrow}$ ($ i\equiv (i_x, i_y)$), where the phase fluctuations in the attractive Hubbard model map to spin fluctuations in the repulsive Hubbard model~\cite{scalettar1989phase}. This connection makes it apparent that with larger $|U|$, one would expect phase fluctuations to get stronger as the superexchange energy $J=-4t^2/U$ gets smaller. 
Throughout this work, we thus compare the behavior of the systems with $U=-10t$ and with $U=-5t$. For $U=-10t$, the monolayer attractive Hubbard model is deep in the \gls*{BEC} regime, with a strong pairing interaction and substantial phase fluctuations, which in turn suppresses $\Tc$. By coupling it to a metal, we hope to mitigate the phase fluctuations and thereby enhance $\Tc$. In contrast, the decoupled monolayer with $U=-5t$ at $\langle n_1\rangle \approx 0.75$ has been found by us to have a $\Tc\approx0.168t$ , which is very close to the optimal values reported previously for the single band $-U$ Hubbard model~\cite{attrHub_BCS-BEC}. Comparing $U=-10t$ and $U=-5t$ results, therefore, provides additional insight into the superconducting character of our bilayer system and the extent to which $\Tc$ can be optimized.  

\subsection{Determinant Quantum Monte Carlo}

We perform \gls*{DQMC} simulations of the Hamiltonian in Eq.~\eqref{eq:hamiltonian} using the \texttt{SmoQyDQMC.jl} package~\cite{smoqy, smoqy_code}. The core idea behind the \gls*{DQMC} algorithm is to perform a Trotter decomposition to the density matrix $e^{\beta \hat{H}}=\Pi_{i=1...L}e^{\Delta \tau \hat{H}}$, where we discretized the imaginary time (the inverse temperature) $\beta$ into $L$ equal steps of size $\Delta \tau = \beta/L$. The choice of a small $\Delta \tau$ allows one to decouple the kinetic and the Hubbard interaction part of the Hamiltonian by virtue of the relation $e^{\beta \hat{H}}=e^{\Delta\tau(\hat{K}+\hat{I})}=e^{\Delta\tau \hat{K}}e^{\Delta\tau\hat{I}}+o(\Delta\tau^2[\hat{K},\hat{I}])$, where $\hat{K}$ and $\hat{I}$ are the kinetic and interaction part of the Hamiltonian respectively. Owing to the relatively strong interactions in this problem, we use a small imaginary-time discretization of $\Delta\tau=0.05$ in the simulations to control the Trotter error. A Hubbard-Stratonovich Transformation is then performed on the interacting part $e^{\Delta\tau\hat I}$ to turn the interaction term that is quartic in Fermionic operator into quadratic, at the expense of introducing auxiliary fields. We obtain estimations of physical quantities based on the Hamiltonian of Eq.~\eqref{eq:hamiltonian} on $N = 2\times L_x \times L_y$ rectangular bilayer clusters with periodic boundary conditions. We generally consider elongated clusters with $L_y \gg L_x$ to obtain reliable estimates for the superfluid density, as discussed in the next section. Most of the data presented in this work were obtained using between $10^5$ and $10^6$ measurement sweeps distributed over $10 - 100$ independent Markov chains. For error analysis, we generated 100 bins of data, and used the Jackknife method to estimate the error bars. 


\subsection{Measurements}  \label{sec: measurements}
Our main goal is to determine the superconducting transition temperature $\Tc$ as a function of the inter-layer hopping $t_\perp$ and examine the evolution of pairing magnitude and phase coherence in the system. We start by defining the $\boldsymbol{q}=0$ $s$-wave pairing structure factor 
\begin{align}
P_{s,ll^\prime}(\tau) = \frac{1}{N}\sum_{i,j} \langle \Delta_{i,l}^{\vphantom{\dagger}}(\tau) \Delta_{j,l^\prime}^{\dagger}(0) \rangle, 
\end{align}
where $i$ indexes the unit cells in the bilayer, $l$ indexes the layer, and $\Delta_{i,l} = \hat{c}_{i,l, \downarrow} \hat{c}_{i,l,\uparrow}$ is a local pair annihilation operator. The corresponding pair-field susceptibility $\mathcal{P}_{s,ll^\prime}$ is obtained by integrating the pairing structure factor over imaginary time 
\begin{align}
\mathcal{P}_{s,ll^\prime} = \int_0^\beta  P_{s,ll^\prime}(\tau)\, d\tau. 
\end{align}
The pairing structure factor $P_{s,ll^\prime}$ gives direct information about whether the Cooper pairs develop long-range correlations in space, while the pair-field susceptibility $\mathcal{P}_{s,ll^\prime}$ diverges at the superconducting instability in the thermodynamic limit. Throughout this work, we also use the shorthand notations $\mathcal{P}_{s,l}\equiv \mathcal{P}_{s,ll}$ and $P_{s,l}\equiv P_{s,ll}$. 

We also measured the superfluid density $\rhos$, which can be obtained from the momentum-resolved current-current correlation function~\cite{SWZ} 
\begin{align}
\frac{\rhos}{\pi e^2} =  -\langle k^\text{tot}_{x} \rangle - \lim_{q_y\rightarrow 0}\Lambda_{xx,}^\text{tot}(q_x=0,q_y,\mathrm{i}\omega_m=0). 
\end{align}
Here, $\langle k^\text{tot}_{x} \rangle$ is the average kinetic energy of the bonds parallel
to the $x$ direction of the bilayer, and 
\begin{equation}
\Lambda^\text{tot}_{xx}(\boldsymbol{q},\mathrm{i}\omega_m) = \frac{1}{N} \int_0^\beta d\tau e^{\mathrm{i}\omega_m\tau} \langle J^{\mrm{p},\text{tot}}_{x}(\boldsymbol{q},\tau) J^{\mrm{p},\text{tot}}_{x}(-\boldsymbol{q},0)\rangle
\end{equation}
with the paramagnetic part of the current operator given by 
\begin{align*}
    J^{\mrm{p},\mathrm{tot}}_{x}(\boldsymbol{q},\tau) = \sum_l J^{\mrm{p}}_{x,l}(\boldsymbol{q},\tau)
\end{align*}
with
\begin{align*}
J^{\mrm{p}}_{x,l}(\boldsymbol{q}) = \mathrm{i}t\sum_{i,\sigma} e^{-\mathrm{i}\boldsymbol{q}\cdot\boldsymbol{R}_i}
\left( \hat{c}_{i,l,\sigma}^\dagger \hat{c}^{\phantom\dagger}_{i+\hat{x},l,\sigma} - \hat{c}^{\vphantom{\boldsymbol{x}+\hat{i}}\dagger}_{i+\hat{x},l,\sigma} \hat{c}_{i,l,\sigma}^{\vphantom{\dagger}} \right).
\end{align*}
In the above expressions, $\omega_m = 2m\pi/\beta$ is a bosonic Matsubara frequency, 
$i$ again runs over all sites in a given layer, $\boldsymbol{R}_i$ is the in-plane position of lattice site $i$, and $i+\hat{x}$ denotes the neighboring site located at $\boldsymbol{R}_i+a\hat{x}$, where $a$ is the in-plane lattice constant. Because the calculation of $\rhos$ involves an extrapolation $q_y \rightarrow 0$, we typically employ rectangular lattices ({\it e.g.}~$L_x \times L_y = 8 \times 32$) to obtain a finer momentum grid in the $q_y$ direction. 

The superfluid density $\rhos$ is a key quantity as it is usually the bottleneck to achieving higher transition temperatures in superconductors with strong interactions~\cite{Kivelson2002, berg2008route} and is directly proportional to the superfluid stiffness $\Ds = \rhos/4\pi e^2$ \cite{attrHub_BCS-BEC}. Importantly, for a \gls*{2D} Kosterlitz-Thouless transition, $\Ds$ has a universal value of $\Ds(T=\Tc) = 2\Tc/\pi$. Thus, we can extract $\Tc$ from the intersection of a straight line of $2T/\pi$ and the $\Ds(T)$ curve as a function of $T$. 

\section{Results}

\subsection{Evolution of the transition temperature with $t_\perp$} \label{sec: PD}

\begin {figure}[t]
\includegraphics[width=0.48\textwidth]{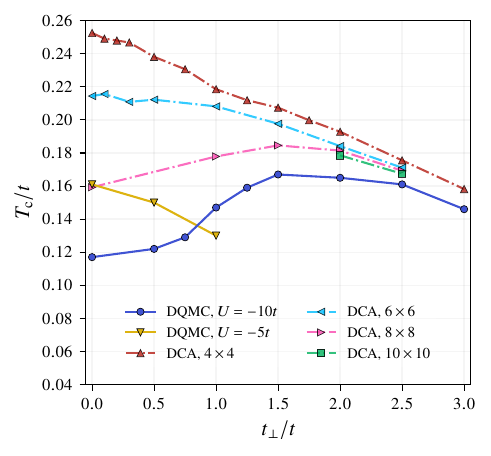}
\caption{$\Tc$ as a function of $t_\perp$ obtained using DQMC for $U=-10t$ and $U=-5t$ bilayers. Previous DCA results (from Ref.~\cite{dee2022enhancing}) are shown for $U=-10t$ and several cluster sizes. 
The results are very close to each other, and $\Tc$ is largest at $t_\perp=1.5$ 
and present the same trend as $t_\perp$ changes for both methods. }
\label{fig:PD}
\end{figure}

\begin{figure*}[t] 
  \centering
  \includegraphics[width=\textwidth]{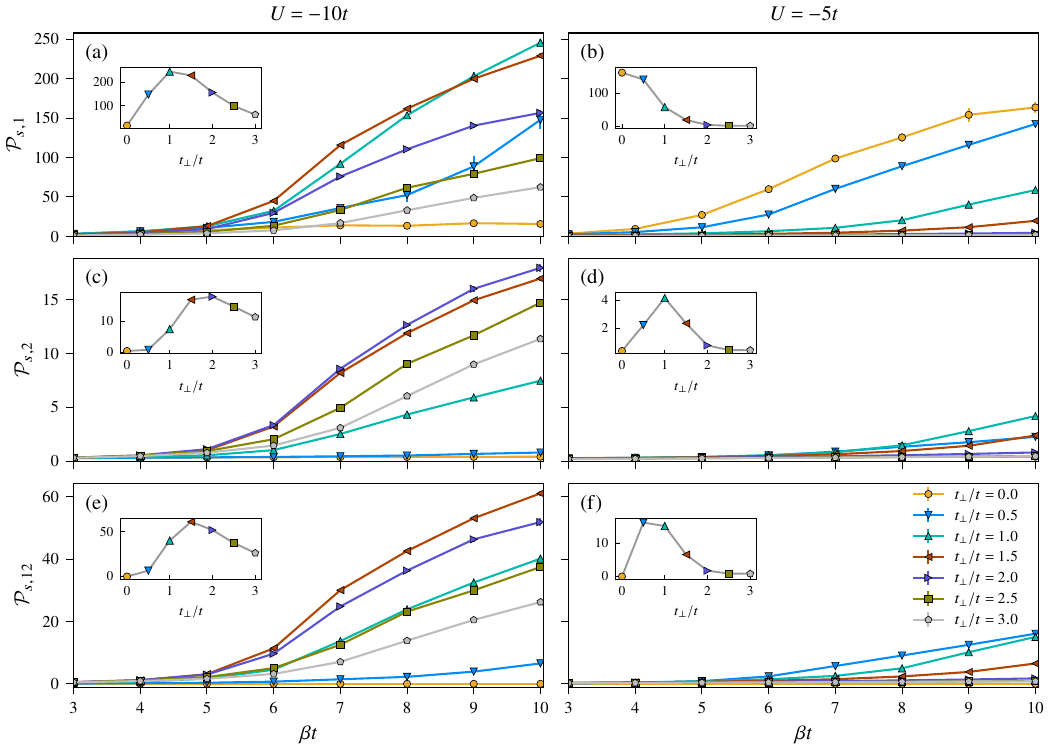} 
  \caption{Layer-resolved pairing susceptibility $\mathcal{P}_{s,ll^\prime}$ on an $8\times 32$ lattice, for $U=-10t$ and $U=-5t$ as functions of inverse temperature $\beta t$. The results for the (a)-(b) correlated layer 1, (c)-(d) metallic layer 2, and the (e)-(f) interlayer region are shown, with corresponding insets depicting $\mathcal{P}_{s,ll^\prime}(\beta t=10)$. Here we use the shorthand notation $\mathcal{P}_{s,l}\equiv \mathcal{P}_{s,ll}$. All lines connecting points are included to guide the eye. }
  \label{fig:Ps_int(beta)}
\end{figure*}

Figure~\ref{fig:PD} presents the evolution of $\Tc$ as a function of $t_\perp$, which is one of our main results. For $U = -10t$, we obtain nonmonotonic behavior with an optimal $\Tc \approx 0.17t$ for $t_\perp = 1.5t$. For reference, we also show the $\Tc$ values obtained from the previous \gls*{DCA} study for different cluster sizes, as indicated in the legend~\cite{dee2022enhancing}. Our \gls*{DQMC} results for $\Tc$ are in general agreement with the \gls*{DCA}~\cite{dee2022enhancing} in that the $\Tc$ estimates obtained from the latter appear to be converging to the former with increasing cluster size. Notably, the rate of convergence is faster in the large $t_\perp/t$ regime. (As discussed in Ref.~\cite{dee2022enhancing}, the \gls*{DCA} calculations exhibit a significant system size dependence for smaller $t_\perp$ values.) As we will show later, the superfluid density, as well as the pair-field susceptibility, show a similar dome-like structure as a function of $t_\perp$ (see Fig.~\ref{fig:Ps_int(beta)}). 

The bilayer system with $U=-5t$ shows qualitatively different behavior, with $\Tc$ decreasing monotonically with increasing $t_\perp$. The pair-field susceptibility and the superfluid density also show similar trends, indicating that $t_\perp$ only suppresses superconductivity in this case. 

The trends of $\Tc$ for different interaction strengths can be understood as follows. 
Strong interactions in the correlated layer generally reduce the superfluid density of that layer. In isolation, as Kivelson pointed out~\cite{Kivelson2002}, the reduced superfluid density suppresses $\Tc$ of the layer well below its mean-field value. (for single attractive Hubbard layer at $U=-10$, $\Tc$ obtained by \gls*{DQMC} is about an order of magnitude lower than the mean-field $\Tc$ by our calculation) Initially coupling the correlated layer to the metal layer thus provides superfluid density to the interacting layer, increasing $\Tc$. 
However, as the inter-layer tunneling $t_\perp$ increases, the layers begin to hybridize strongly, forming bonding and antibonding states. In the limit $t_\perp \gg t$, the electrons tend to condense into local dimer states formed from the bonding combination of the orbitals in each layer. This dimerization results in a smaller effective interaction between the electrons and, thus, a smaller mean-field $\Tc$ value for the system as a whole. Consequently, it is initially beneficial to couple the strong interacting layer with a metal, but eventually the decrease in the effective interaction takes over and suppresses $\Tc$.  
For the $U=-5t$ system, the correlated layer has already been optimized with respect to $\Tc$ and is less affected by phase fluctuations. Coupling the correlated layer to the metallic layer thus has no benefit and only reduces $\Tc$ via the reduction in the overall effective interaction.

In the following sections, we will provide a more detailed analysis of the measurements leading to Fig.~\ref{fig:PD} and the physical picture we have just described.

\subsection{Pair-field Susceptibility} \label{sec: Ps}
Figure~\ref{fig:Ps_int(beta)} plots the pair-field susceptibility $\mathcal{P}_{s,ll^{\prime}}$ as a function of temperature for $U = -10t$ (left column) and $-5t$ (right column). A divergence of the pair-field susceptibility in the thermodynamic limit would imply a superconducting instability. However, these divergences are cut off by the finite size of the cluster in our calculations. Each column in Fig.~\ref{fig:Ps_int(beta)} shows results for the interacting ($\mathcal{P}_{s,11}$, top row) and metallic ($\mathcal{P}_{s,22}$, middle row) layers, as well as the inter-layer ($\mathcal{P}_{s,12}$, bottom row) pair-field susceptibility for different $t_\perp \in [0,3t]$, as indicated by the common legend provided in Fig.~\ref{fig:Ps_int(beta)}f. The insets of each panel show the dependence of the pair-field susceptibility at the largest $\beta =10/t$ as a function of $t_\perp$. 

We first focus on the $U = -10t$ case. Fig.~\ref{fig:Ps_int(beta)}a shows that the pair-field susceptibility for the correlated layer $\mathcal{P}_{s,1}$ generally increases as the temperature is lowered, and in some cases, shows signs of leveling off at low-temperatures, depending on the value of $t_\perp$. The inset of Fig.~\ref{fig:Ps_int(beta)}a summarizes the evolution of $\mathcal{P}_{s,1}$ 
at $\beta =10/t$ as a function of $t_\perp$; it has a nonmonotonic behavior, first increasing and then decreasing as $t_\perp$ grows from zero. In fact, $\mathcal{P}_{s,1}$ is enhanced by more than an order of magnitude at this temperature, growing from 15 at $t_\perp=0$ (the single-layer limit) to the highest value of 245 with $t_\perp=t$. This enhancement is linked to extended correlation lengths of the pairing correlations in both spatial and temporal directions, as discussed further in App.~\ref{Appendix2}. We conclude that the coupling of a $U=-10t$ interacting layer to a metal layer with moderate $t_\perp$ value can strongly boost the pair-field susceptibility of the interacting layer by enhancing its superfluid density (stiffness) and allowing the phases to be coherent over longer spatial and temporal distances. However, $\mathcal{P}_{s,1}$ begins to drop for $t_\perp>1.5t$; with a continued increase in $t_\perp$, the bandwidth $W$ of the underlying noninteracting part of the Hamiltonian increases further, which brings down the value of $U/W$ and eventually pushes the system toward the noninteracting limit, suppressing superconductivity. 

In contrast to the strong coupling case, $\mathcal{P}_{s,1}$ for intermediate coupling $U=-5t$, shown in Fig.~\ref{fig:Ps_int(beta)}b, exhibits a monotonic decrease as $t_\perp$ is introduced. This behavior can be observed in both the temperature dependence of the data, as well as the $t_\perp$ dependence at $\beta = 10/t$ shown in the inset. Since $U=-5t$ with a filling of $\langle n_1\rangle=0.75$ is in the parameter range where $\Tc$ is optimized for a single layer \cite{attrHub_BCS-BEC}, the introduction of $t_\perp$ primarily encourages the system to form bonding and anti-bonding states between the layers. This hybridization reduces the system's effective interaction and thus drives the system away from the optimal $\Tc$ obtained at $t_\perp = 0$. It is noteworthy that the absolute value of the largest $\mathcal{P}_{s,1}$ for $U=-10t$ (at $t_\perp=1.5t$ and $\beta =10/t$) is significantly larger than that for $U=-5t$ at $t_\perp=0$ (single layer limit) and $\beta =10/t$. This difference likely reflects an increase in the superconducting gap in the composite system with $U=-10t$ compared to the optimized single layer with $U=-5t$. Since $U=-5t$ has the highest possible $\Tc$ for a single layer, this result suggests that the effect of $t_\perp$ on the $U=-10t$ layer is not merely decreasing its effective interaction and bringing it to a more weakly interacting regime. A boost in superfluid density is also observed in the bilayer system over that of the optimal single layer and will be discussed in Sec.~\ref{sec: curr}. These observations strongly imply that the bilayer system cannot be simply mapped onto an effective singleband model with a lower $U_\mathrm{eff}$ derived from the bonding orbitals of the bilayer. 

Examining the pair-field susceptibility $\mathcal{P}_{s,2}$ of the metal layer tells us more about how strong the proximity effect is. Comparing the overall scales of the top and middle rows of Fig.~\ref{fig:Ps_int(beta)}, we see that $\mathcal{P}_{s,2}$ is always much smaller in magnitude than $\mathcal{P}_{s,1}$ for both values of the Hubbard interaction. From the insets of Fig.~\ref{fig:Ps_int(beta)}c and Fig.~\ref{fig:Ps_int(beta)}d, we can see that $\mathcal{P}_{s,2}$ in both systems also shows a nonmonotonic dependence on $t_\perp$. Similar to $\mathcal{P}_{s,2}$, in the inset of panel Fig.~\ref{fig:Ps_int(beta)}e and Fig.~\ref{fig:Ps_int(beta)}f, the behavior of $\mathcal{P}_{s,12}$ closely resembles that of $\mathcal{P}_{s,2}$, but is stronger in magnitude. The inter-layer correlation is a convolution of the pairing strength of the two layers, and has a magnitude intermediate between the two.
These observations are fully consistent with what one might expect for pairing in the metallic layer, which is induced by the proximity effect.

\subsection{Current Correlation and Superfluid Density} \label{sec: curr}

\begin{figure}[t] 
  \centering
  \includegraphics[width=\columnwidth]{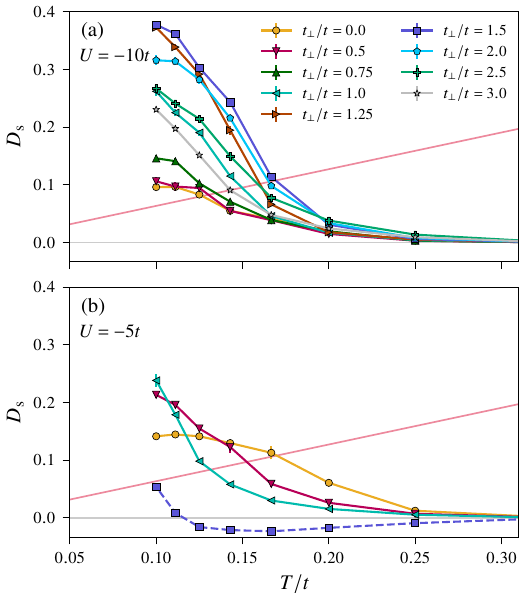} 
  \caption{Total superfluid stiffness $\Ds$ on and $8\times 32$ lattice, for (a) $U=-10t$ (b) and $U=-5t$ as functions of temperature $T/t$.  All lines connecting points are included to guide the eye. A dashed line is used for the $t_{\perp}=1.5t$ data in the $U=-5t$ case to indicate that estimations for $\Ds$ became unreliable for $t_{\perp}\geq 1.5t$. }
  \label{fig:D_s}
\end{figure}

We now turn to the superfluid density $\Ds$ as a function of the temperature $T$. Figs.~\ref{fig:D_s}a and \ref{fig:D_s}b show how $\Ds$ evolves for different $t_\perp$ for $U=-10t$ and $-5t$, respectively. The superfluid density, proportional to the helicity modulus, is known to have a universal value of $2\Tc/\pi$ at the critical temperature for systems belonging to the 2D XY universality 
class.\footnote{This universality class can be justified by considering the combined gauge and partial particle-hole transformation $c_{i\downarrow}^{\phantom\dagger} \leftrightarrow (-1)^i c_{i\downarrow}^{\dagger}$, which maps the attractive Hubbard model with a nonzero chemical potential $\mu (n_{\uparrow} + n_{\downarrow})$ onto the repulsive Hubbard model with a Zeeman field $\mu (n_{\uparrow } - n_{\downarrow})$. This Zeeman field breaks the Heisenberg symmetry of the antiferromagnetic correlations in the repulsive Hubbard model.  It is energetically favorable for the spins to `lie down' in the $xy$-plane perpendicular to the $z$ direction~\cite{scalettar1989phase}. 
} 
As such, the critical temperature $\Tc$ of the system can be extracted for a particular value of $t_\perp$ from the intersection of a straight line with slope $2T/\pi$ and the $\Ds(T)$ curves obtained from our \gls*{DQMC} calculations. In App.~\ref{Appendi_finite_size}, we present a comparison of $D_s$ evaluated on $6\times32$ and $8\times32$ systems for the $U=-10\,t$ bilayer to demonstrate that finite size effects are well controlled for the systems considered here.

For $U = -10t$, shown in Fig.~\ref{fig:D_s}a, the magnitude of $\Ds$ as well as the intersection of it with the $2T/\pi$ line ($\Tc$) show the same nonmonotonic trend of increase followed by a decrease as a function of $t_\perp$. This behavior is consistent with the trends in the pair-field susceptibilities shown in Fig.~\ref{fig:Ps_int(beta)}. In this case, $\Tc$ reaches a maximum of $\approx0.17t$ for $t_\perp=1.5t$, again in agreement with the behavior of the pair-field susceptibility at low temperatures. 

Note that the $\Ds(T)$ curve for $t_\perp=0.5t$ shows a very small enhancement over that of $t_\perp=0$, which aligns more closely with the trend of $\mathcal{P}_{s,2}$ of the metal layer, shown in the inset of Fig.~\ref{fig:Ps_int(beta)}c. At the same time, $\mathcal{P}_{s,1}$ (Fig.~\ref{fig:Ps_int(beta)}a) at $t_\perp=0.5t$ already shows substantial growth over the $t_\perp=0$ value, which however, doesn't help too much in increasing $\Ds$. These results are consistent with the proposal that the boost in total superfluid stiffness is provided by pairs residing in the metal layer~\cite{Kivelson2002, berg2008route, wachtel2012superfluid}. 

For $U=-5t$, shown in Fig.~\ref{fig:D_s}b, we see a monotonic decrease of $\Ds$ in the high-temperature region as $t_\perp$ increases. $\Tc$, obtained by the crossing points with the $2T/\pi$ line when it occurs, thus also decreases monotonically. This trend also agrees with the pair-field susceptibility behavior shown in Fig.~\ref{fig:Ps_int(beta)}d. However, at temperatures lower than $\Tc$, $\Ds$ for the $t_\perp=0.5t$ and $t$ is enhanced over the value obtained for an isolated layer ($t_\perp=0$). Thus, coupling the metal layer to the interacting one can enhance the superfluid density of the interacting layer, even for model parameters ($U=-5t$ and $\langle n \rangle=0.75$) that maximize $\Tc$ when the correlated layer is isolated. The enhanced superfluid density only occurs for $T<\Tc$, however, and may not always act to increase $\Tc$. Importantly, comparing the magnitude of $\Ds$ of $U=-10$ in Figs.~\ref{fig:D_s}a and $U=-5$ in \ref{fig:D_s}b, we find that the maximum $\Ds$ that can be obtained in the bilayer system with $U=-10t$ at $T=0.1t$ is much higher than the values obtained for $U=-5t$ at any $t_\perp$, including $t_\perp=0$, which corresponds to the optimally tuned monolayer. This mirrors the behavior of $\mathcal{P}_{s,1}$ discussed in Sec.~\ref{sec: Ps}.

Lastly, we have included data for $t_{\perp}=1.5t$ in Fig.~\ref{fig:D_s}b using a dashed guideline to highlight this result. For this value of $t_\perp$, we encountered numerical challenges extracting quality estimates of $\Ds$, even for the elongated $8\times 32$ lattice. While the eventual intersection of this curve with the $2T/\pi$ line indeed yields a $\Tc$ that continues to follow the monotonic decrease with $t_{\perp}$, it also displays negative stiffness values at several temperatures above the putative $\Tc$. We believe that for these parameters of $U$ and large $t_\perp$, where the effective interaction and therefore the pairing amplitude become small, finite-size effects become substantial, making accurate estimates difficult for the lattice size we have used.  

\section{Summary \& Discussion} \label{sec: conc}
We have examined how the pair-field susceptibility $\mathcal{P}_s$ and superfluid density $\Ds$ evolve in a bilayer system comprised of an attractive Hubbard layer coupled to a noninteracting metallic layer as a function of the interlayer tunneling  $t_\perp$. By exploiting the universal value of $\Ds(\Tc) = 2\Tc/\pi$ for a 2D superconductor, we also obtained the $\Tc$ for different values of $t_\perp$. The non-monotonic behavior of $\Tc$ for $U = -10t$ aligns with the findings of a prior \gls*{DCA} study~\cite{dee2022enhancing}, despite the significant finite size effects observed in that case for small $t_\perp$. Our work also extends the work reported in Refs.~\cite{berg2008route} and \cite{wachtel2012superfluid} by including a non-zero intralayer hopping in the correlated layer as opposed to considering independent $-U$ sites (i.e., a periodic Anderson impurity model).

We have also compared the behavior of the bilayer with $U = -10t$ and $U=-5t$ to obtain additional physical insights. For $U = -5t$ and the density we simulated for the correlated layer, the \gls*{2D} $-U$ Hubbard model has an optimized $\Tc$. We found that coupling such an optimized layer with a metal layer results in a monotonic decrease in $\Tc$ as the interlayer hopping increases. This behavior can be rationalized as resulting from an overall decrease in the effective interaction acting in the system, which drives it into a sub-optimal weakly-interacting regime. However, we also found some evidence that this is not the only effect that takes place. Even though Fig.~\ref{fig:PD} shows that the highest $\Tc$ achieved in the $U=-10t$ bilayer is not notably higher than the optimal $\Tc$ for the \gls*{2D} singleband $-U$ model, the enhanced pair-field susceptibility (i.e., gap size) and superfluid density of the composite system are significantly larger. In other words, while the $\Tc$ gains may be relatively modest in the composite system, the condensate itself may be substantially more robust, which has implications for technological applications. 

We also showed that $\Ds$ increases with $t_\perp$, in agreement with Kivelson's original proposal~\cite{Kivelson2002}. Moreover, this increase is always accompanied by a substantial enhancement of the metallic layer's pair-field susceptibility 
$\mathcal{P}_{s,2}$. However, the enhancement of $\Ds$ may only occur for $T < \Tc$ in some cases, as seen in Fig.~\ref{fig:D_s}b, so it doesn't necessarily drive $\Tc$ up. We believe that a reduction in effective interaction and a boost in superfluid stiffness happen simultaneously and are the two main drivers of changes to $\Tc$ for the bilayer. Their interplay is responsible for the rich physics of the composite system.

Finally, the parameter space of the composite system is large. Here we have focused on a specific choice of $t_l$ and $\epsilon_l$, and tuned $\mu$ to fix the filling of the correlated layer ($n_{1}=0.75$). However, other unexplored parameter regimes are particularly interesting, including where the Fermi level matches the van Hove singularity or when the Fermi surfaces of the two layers coincide. Other fillings, on-site energy differences, and hopping strengths of both layers are interesting parameters to look at as well, though the case $t_1=0$ (a periodic Anderson model with $-U$ centers) was the focus of Refs.~\cite{berg2008route} and \cite{wachtel2012superfluid}. Ref.~\cite{wachtel2012superfluid} also explored the interesting case when chemical potential disorder is included in the metallic layer, allowing for a discussion of the insensitivity of pairing to randomness in a case when the attractive sites are distinct from the location of the disorder. Another line of future inquiry is to explore the effect of weak attractive interaction or retarded interaction induced by phonons in the metal layer. In that case, pairing in the metal layers will not be purely induced by the proximity effect, which could further increase $\Tc$ and introduce additional competing charge order instabilities or other polaronic effects. 

\section*{Acknowledgments} This work was supported by the  U.S.~Department of Energy, Office of Science, Office of Basic Energy Sciences, under Award Number DE-SC0022311. \\


\appendix

\renewcommand{\thefigure}{A\arabic{figure}}
\setcounter{figure}{0}

\section{Imaginary time pair-field susceptibility and equal-time pairing structure factor} \label{Appendix2}

\begin{figure*}[t]
\includegraphics[width=\textwidth]{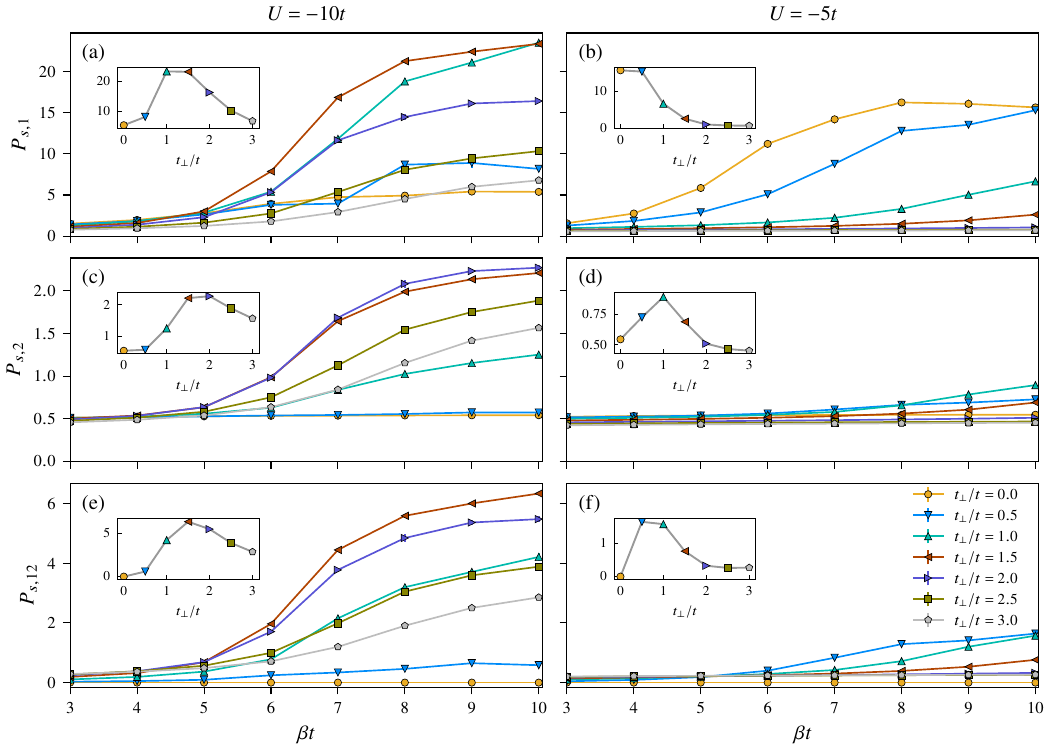}
\caption{Layer-resolved equal-time pairing structure factor $P_{s,ll^{\prime}}$ on an $8\times 32$ lattice, for $U=-10t$ and $U=-5t$ as functions of inverse temperature $\beta t$. The results for the (a)-(b) correlated layer 1, (c)-(d) metallic layer 2, and the (e)-(f) interlayer region are shown, with corresponding insets depicting $P_{s,ll^{\prime}}(\beta t=10)$. Here, we use the shorthand notation $P_{s,l}\equiv P_{s,ll}$. All lines connecting points are included to guide the eye. }
\label{fig:Ps_eqt}
\end{figure*}

\begin{figure}[ht]
\includegraphics[width=\columnwidth]{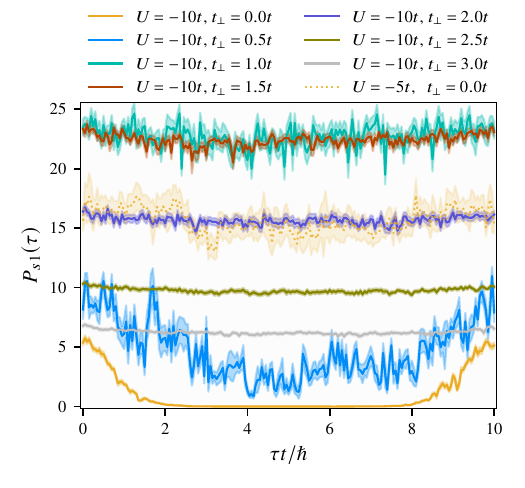}
\caption{Pairing structure factor $P_{s,1}(\tau)$ on an $8\times32$ lattice, as a function of the imaginary time $\tau$, for different $t_\perp$ values. Solid lines show $P_{s,1}$ for $U_1=-10t$ at different $t_\perp$ values. Dashed line show $P_{s,1}$ for $U_1=-5t$ at $t_\perp=0.0$ (single-layer) as a comparison. It can be seen that with intermediate $t_\perp$ values, not only does the equal-time $P_{s,1}(\tau=0)$ get increased, but $P_{s,1}(\tau)$ also becomes longer ranged in imaginary time. $P_{s,1}$ for $U_1=-5t$ single layer also shows a slow decay in imaginary time, which resembles that of $U_1=-10t$ at $t_\perp=2t$.}
\label{fig:Ps(tau)}
\end{figure}

Here, we provide additional results for the equal-time structure factor $P_{s,ll^{\prime}}$ and the imaginary-time dependence of the pairing structure factor $P_{s,1}(\tau)$.

Figure~\ref{fig:Ps_eqt} plots the equal-time pairing structure factor $P_{s,ll^{\prime}}$. The results are similar to the pair-field susceptibility $\mathcal{P}_{s,ll^{\prime}}$. Note that $P_{s,ll^{\prime}}$ directly measures the real-space pair correlations. If we look at the inset of Fig.~\ref{fig:Ps_eqt}a, $P_{s,1}$ is increased by about a factor of five when $t_\perp$ is increased from $0.0$ to $1.5t$.  While this is a smaller increase than that of $\mathcal{P}_{s,ll^{\prime}}$ (a factor or about $16$ times for the same parameters), it is nonetheless significant.

The origin of the extra boost in $\mathcal{P}_{s,ll^{\prime}}$ can be explained by the imaginary time dependence of $P_{s,1}(\tau)$ shown in  Fig.~\ref{fig:Ps(tau)}. For $U=-10t$ (solid lines), comparing the curve of $t_\perp=0.0$ with the curves for non-zero $t_\perp$ reveals a drastic change in the long-range correlation in $\tau$. The $P_{s,1}(\tau)$ curve at $t_\perp=0.0$ exhibits a clear exponential decay with $\tau$. When $t_\perp$ is increased to $0.5t$, $P_{s,1}(\tau)$ begins to show quasi-long-range correlations but with significant fluctuations and a decaying structure in $\tau$. However, for $t_\perp>0.5t$, $P_{s,1}(\tau)$ becomes nearly independent of $\tau$ and with much smaller fluctuations. Since $\mathcal{P}_{s,ll^{\prime}}$ is set by the area under the $P_{s,1}(\tau)$ curve, the growth in $\mathcal{P}_{s,ll^{\prime}}$ can be traced back to the increased correlation length not only in space but also in imaginary time. 

For comparison, a $U=-5t$ curve (dashed line) in the single-layer limit ($t_\perp=0.0$) is plotted. Interestingly, the $U=-5t$ curve matches pretty well with the $U=-10t$ curves at $t_\perp=2t$ but resides below the $U=-10t$, $t_\perp=t$ and $1.5t$ curves. Since $U=-5t$ represents the system with the highest $T_c$ for a single layer, this observation further emphasizes that the effect of $t_\perp$ is not merely reducing the effective interaction -- the composite system can have a larger $P_{s,1}(\tau)$ compared to an optimized single layer $-U$ model. On the other hand, the change of the shape in $P_{s,1}(\tau)$ indicates that the effect of lowered effective interaction is also at play, as the change of the $\tau$-dependence of $P_{s,1}(\tau)$ from $t_\perp=0.0$ to $t_\perp>0$ (for $U=-10t$) resembles that from $U=-10t$ single layer to $U=-5t$ single layer. Thus, 
the reduction in effective interaction, which reduces phase fluctuations, is one of the driving forces enhancing $\Tc$, in addition to the increase in the superfluid density described in the main text.

\section{Superfluid Density Evaluated at Different System Sizes} \label{Appendi_finite_size}

\begin{figure}[ht] 
  \centering
  \includegraphics[width=\columnwidth]{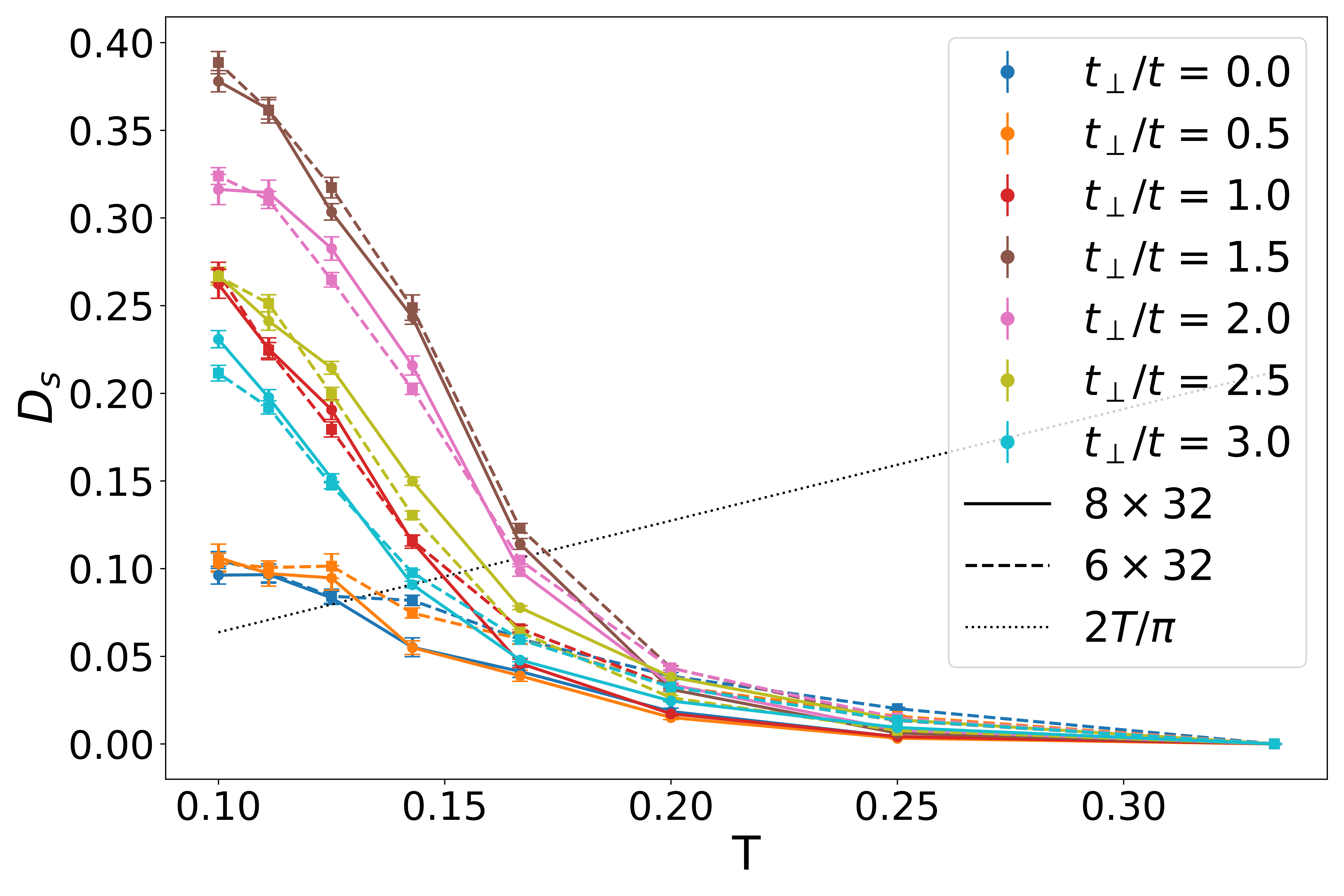} 
  \caption{Total superfluid stiffness $D_s$ on $6\times 32$ and $8\times 32$ lattice, for $U=-10t$ as functions of temperature $T/t$.  All lines connecting points are included to guide the eye. A dashed line is used for the $6\times 32$ lattice data.}
  \label{fig:finite_size}
\end{figure}

Here we compare the total superfluid density $D_s$ of the $U=-10\,t$ system to determine whether finite size effects are under control. Fig.~\ref{fig:finite_size} plots the intersections of the $D_s(T)$ curves with $2T/\pi$ 
for both $6\times32$ and $8\times32$ lattices. The extracted $\Tc$ values for both system sizes are very similar, indicating that the finite-size effects are small and within statistical error bars from the Monte Carlo sampling. Most importantly, the trend of how $\Tc$ changes as $t_\perp$ increases is consistent across the two system sizes. 

\bibliography{references}

\begin{thebibliography}{24}%
\makeatletter
\providecommand \@ifxundefined [1]{%
 \@ifx{#1\undefined}
}%
\providecommand \@ifnum [1]{%
 \ifnum #1\expandafter \@firstoftwo
 \else \expandafter \@secondoftwo
 \fi
}%
\providecommand \@ifx [1]{%
 \ifx #1\expandafter \@firstoftwo
 \else \expandafter \@secondoftwo
 \fi
}%
\providecommand \natexlab [1]{#1}%
\providecommand \enquote  [1]{``#1''}%
\providecommand \bibnamefont  [1]{#1}%
\providecommand \bibfnamefont [1]{#1}%
\providecommand \citenamefont [1]{#1}%
\providecommand \href@noop [0]{\@secondoftwo}%
\providecommand \href [0]{\begingroup \@sanitize@url \@href}%
\providecommand \@href[1]{\@@startlink{#1}\@@href}%
\providecommand \@@href[1]{\endgroup#1\@@endlink}%
\providecommand \@sanitize@url [0]{\catcode `\\12\catcode `\$12\catcode `\&12\catcode `\#12\catcode `\^12\catcode `\_12\catcode `\%12\relax}%
\providecommand \@@startlink[1]{}%
\providecommand \@@endlink[0]{}%
\providecommand \url  [0]{\begingroup\@sanitize@url \@url }%
\providecommand \@url [1]{\endgroup\@href {#1}{\urlprefix }}%
\providecommand \urlprefix  [0]{URL }%
\providecommand \Eprint [0]{\href }%
\providecommand \doibase [0]{https://doi.org/}%
\providecommand \selectlanguage [0]{\@gobble}%
\providecommand \bibinfo  [0]{\@secondoftwo}%
\providecommand \bibfield  [0]{\@secondoftwo}%
\providecommand \translation [1]{[#1]}%
\providecommand \BibitemOpen [0]{}%
\providecommand \bibitemStop [0]{}%
\providecommand \bibitemNoStop [0]{.\EOS\space}%
\providecommand \EOS [0]{\spacefactor3000\relax}%
\providecommand \BibitemShut  [1]{\csname bibitem#1\endcsname}%
\let\auto@bib@innerbib\@empty
\bibitem [{\citenamefont {Tsymbal}\ \emph {et~al.}(2012)\citenamefont {Tsymbal}, \citenamefont {Dagotto}, \citenamefont {Eom},\ and\ \citenamefont {Ramesh}}]{Tsymbal2012multifunctional}%
  \BibitemOpen
  \bibfield  {author} {\bibinfo {author} {\bibfnamefont {E.~Y.}\ \bibnamefont {Tsymbal}}, \bibinfo {author} {\bibfnamefont {E.~R.~A.}\ \bibnamefont {Dagotto}}, \bibinfo {author} {\bibfnamefont {C.-B.}\ \bibnamefont {Eom}},\ and\ \bibinfo {author} {\bibfnamefont {R.}~\bibnamefont {Ramesh}},\ }\href {https://doi.org/10.1093/acprof:oso/9780199584123.001.0001} {\emph {\bibinfo {title} {Multifunctional Oxide Heterostructures}}}\ (\bibinfo  {publisher} {Oxford University Press},\ \bibinfo {year} {2012})\BibitemShut {NoStop}%
\bibitem [{\citenamefont {Shepelin}\ \emph {et~al.}(2023)\citenamefont {Shepelin}, \citenamefont {Tehrani}, \citenamefont {Ohannessian}, \citenamefont {Schneider}, \citenamefont {Pergolesi},\ and\ \citenamefont {Lippert}}]{Shepelin2023practicalguide}%
  \BibitemOpen
  \bibfield  {author} {\bibinfo {author} {\bibfnamefont {N.~A.}\ \bibnamefont {Shepelin}}, \bibinfo {author} {\bibfnamefont {Z.~P.}\ \bibnamefont {Tehrani}}, \bibinfo {author} {\bibfnamefont {N.}~\bibnamefont {Ohannessian}}, \bibinfo {author} {\bibfnamefont {C.~W.}\ \bibnamefont {Schneider}}, \bibinfo {author} {\bibfnamefont {D.}~\bibnamefont {Pergolesi}},\ and\ \bibinfo {author} {\bibfnamefont {T.}~\bibnamefont {Lippert}},\ }\bibfield  {title} {\bibinfo {title} {A practical guide to pulsed laser deposition},\ }\href {https://doi.org/10.1039/D2CS00938B} {\bibfield  {journal} {\bibinfo  {journal} {Chem. Soc. Rev.}\ }\textbf {\bibinfo {volume} {52}},\ \bibinfo {pages} {2294} (\bibinfo {year} {2023})}\BibitemShut {NoStop}%
\bibitem [{\citenamefont {Ha}\ and\ \citenamefont {Chung}(2024)}]{Ha2024thinfilm}%
  \BibitemOpen
  \bibfield  {author} {\bibinfo {author} {\bibfnamefont {C.}~\bibnamefont {Ha}}\ and\ \bibinfo {author} {\bibfnamefont {Y.~J.}\ \bibnamefont {Chung}},\ }\bibfield  {title} {\bibinfo {title} {Thin films as practical quantum materials: A status quo and beyond},\ }\href {https://doi.org/10.1063/5.0235472} {\bibfield  {journal} {\bibinfo  {journal} {APL Materials}\ }\textbf {\bibinfo {volume} {12}},\ \bibinfo {pages} {120901} (\bibinfo {year} {2024})}\BibitemShut {NoStop}%
\bibitem [{\citenamefont {Wang}\ \emph {et~al.}(2012)\citenamefont {Wang}, \citenamefont {Li}, \citenamefont {Zhang}, \citenamefont {Zhang}, \citenamefont {Zhang}, \citenamefont {Li}, \citenamefont {Hao}, \citenamefont {Ou}, \citenamefont {Deng}, \citenamefont {Chang}, \citenamefont {Wen}, \citenamefont {Song}, \citenamefont {He}, \citenamefont {Jia}, \citenamefont {Ji}, \citenamefont {Wang}, \citenamefont {Wang}, \citenamefont {Chen}, \citenamefont {Ma},\ and\ \citenamefont {Xue}}]{Wang2012interface}%
  \BibitemOpen
  \bibfield  {author} {\bibinfo {author} {\bibfnamefont {Q.-Y.}\ \bibnamefont {Wang}}, \bibinfo {author} {\bibfnamefont {Z.}~\bibnamefont {Li}}, \bibinfo {author} {\bibfnamefont {W.-H.}\ \bibnamefont {Zhang}}, \bibinfo {author} {\bibfnamefont {Z.-C.}\ \bibnamefont {Zhang}}, \bibinfo {author} {\bibfnamefont {J.-S.}\ \bibnamefont {Zhang}}, \bibinfo {author} {\bibfnamefont {W.}~\bibnamefont {Li}}, \bibinfo {author} {\bibfnamefont {D.}~\bibnamefont {Hao}}, \bibinfo {author} {\bibfnamefont {Y.-B.}\ \bibnamefont {Ou}}, \bibinfo {author} {\bibfnamefont {P.}~\bibnamefont {Deng}}, \bibinfo {author} {\bibfnamefont {K.}~\bibnamefont {Chang}}, \bibinfo {author} {\bibfnamefont {J.}~\bibnamefont {Wen}}, \bibinfo {author} {\bibfnamefont {C.-L.}\ \bibnamefont {Song}}, \bibinfo {author} {\bibfnamefont {K.}~\bibnamefont {He}}, \bibinfo {author} {\bibfnamefont {J.-F.}\ \bibnamefont {Jia}}, \bibinfo {author} {\bibfnamefont {S.-H.}\ \bibnamefont {Ji}}, \bibinfo {author} {\bibfnamefont {Y.-Y.}\ \bibnamefont {Wang}}, \bibinfo
  {author} {\bibfnamefont {L.-L.}\ \bibnamefont {Wang}}, \bibinfo {author} {\bibfnamefont {X.}~\bibnamefont {Chen}}, \bibinfo {author} {\bibfnamefont {X.-C.}\ \bibnamefont {Ma}},\ and\ \bibinfo {author} {\bibfnamefont {Q.-K.}\ \bibnamefont {Xue}},\ }\bibfield  {title} {\bibinfo {title} {Interface-induced high-temperature superconductivity in single unit-cell {FeSe} films on {SrTiO$_3$}},\ }\href {https://doi.org/10.1088/0256-307X/29/3/037402} {\bibfield  {journal} {\bibinfo  {journal} {Chinese Physics Letters}\ }\textbf {\bibinfo {volume} {29}},\ \bibinfo {pages} {037402} (\bibinfo {year} {2012})}\BibitemShut {NoStop}%
\bibitem [{\citenamefont {Bozovic}\ and\ \citenamefont {Ahn}(2014)}]{Bozovic2014new}%
  \BibitemOpen
  \bibfield  {author} {\bibinfo {author} {\bibfnamefont {I.}~\bibnamefont {Bozovic}}\ and\ \bibinfo {author} {\bibfnamefont {C.}~\bibnamefont {Ahn}},\ }\bibfield  {title} {\bibinfo {title} {A new frontier for superconductivity},\ }\href {https://doi.org/10.1038/nphys3177} {\bibfield  {journal} {\bibinfo  {journal} {Nature Physics}\ }\textbf {\bibinfo {volume} {10}},\ \bibinfo {pages} {892} (\bibinfo {year} {2014})}\BibitemShut {NoStop}%
\bibitem [{\citenamefont {Huang}\ and\ \citenamefont {Hoffman}(2017)}]{Huang2017monolayer}%
  \BibitemOpen
  \bibfield  {author} {\bibinfo {author} {\bibfnamefont {D.}~\bibnamefont {Huang}}\ and\ \bibinfo {author} {\bibfnamefont {J.~E.}\ \bibnamefont {Hoffman}},\ }\bibfield  {title} {\bibinfo {title} {Monolayer {FeSe} on {SrTiO$_3$}},\ }\href {https://doi.org/https://doi.org/10.1146/annurev-conmatphys-031016-025242} {\bibfield  {journal} {\bibinfo  {journal} {Annual Review of Condensed Matter Physics}\ }\textbf {\bibinfo {volume} {8}},\ \bibinfo {pages} {311} (\bibinfo {year} {2017})}\BibitemShut {NoStop}%
\bibitem [{\citenamefont {Wu}\ \emph {et~al.}(2020)\citenamefont {Wu}, \citenamefont {Ming}, \citenamefont {Smith}, \citenamefont {Liu}, \citenamefont {Ye}, \citenamefont {Wang}, \citenamefont {Johnston},\ and\ \citenamefont {Weitering}}]{Wu2020superconductivity}%
  \BibitemOpen
  \bibfield  {author} {\bibinfo {author} {\bibfnamefont {X.}~\bibnamefont {Wu}}, \bibinfo {author} {\bibfnamefont {F.}~\bibnamefont {Ming}}, \bibinfo {author} {\bibfnamefont {T.~S.}\ \bibnamefont {Smith}}, \bibinfo {author} {\bibfnamefont {G.}~\bibnamefont {Liu}}, \bibinfo {author} {\bibfnamefont {F.}~\bibnamefont {Ye}}, \bibinfo {author} {\bibfnamefont {K.}~\bibnamefont {Wang}}, \bibinfo {author} {\bibfnamefont {S.}~\bibnamefont {Johnston}},\ and\ \bibinfo {author} {\bibfnamefont {H.~H.}\ \bibnamefont {Weitering}},\ }\bibfield  {title} {\bibinfo {title} {Superconductivity in a hole-doped {M}ott-insulating triangular adatom layer on a silicon surface},\ }\href {https://doi.org/10.1103/PhysRevLett.125.117001} {\bibfield  {journal} {\bibinfo  {journal} {Phys. Rev. Lett.}\ }\textbf {\bibinfo {volume} {125}},\ \bibinfo {pages} {117001} (\bibinfo {year} {2020})}\BibitemShut {NoStop}%
\bibitem [{\citenamefont {Ming}\ \emph {et~al.}(2023)\citenamefont {Ming}, \citenamefont {Wu}, \citenamefont {Chen}, \citenamefont {Wang}, \citenamefont {Mai}, \citenamefont {Maier}, \citenamefont {Strockoz}, \citenamefont {Venderbos}, \citenamefont {Gonz{\'a}lez}, \citenamefont {Ortega}, \citenamefont {Johnston},\ and\ \citenamefont {Weitering}}]{Ming2023evidence}%
  \BibitemOpen
  \bibfield  {author} {\bibinfo {author} {\bibfnamefont {F.}~\bibnamefont {Ming}}, \bibinfo {author} {\bibfnamefont {X.}~\bibnamefont {Wu}}, \bibinfo {author} {\bibfnamefont {C.}~\bibnamefont {Chen}}, \bibinfo {author} {\bibfnamefont {K.~D.}\ \bibnamefont {Wang}}, \bibinfo {author} {\bibfnamefont {P.}~\bibnamefont {Mai}}, \bibinfo {author} {\bibfnamefont {T.~A.}\ \bibnamefont {Maier}}, \bibinfo {author} {\bibfnamefont {J.}~\bibnamefont {Strockoz}}, \bibinfo {author} {\bibfnamefont {J.~W.~F.}\ \bibnamefont {Venderbos}}, \bibinfo {author} {\bibfnamefont {C.}~\bibnamefont {Gonz{\'a}lez}}, \bibinfo {author} {\bibfnamefont {J.}~\bibnamefont {Ortega}}, \bibinfo {author} {\bibfnamefont {S.}~\bibnamefont {Johnston}},\ and\ \bibinfo {author} {\bibfnamefont {H.~H.}\ \bibnamefont {Weitering}},\ }\bibfield  {title} {\bibinfo {title} {Evidence for chiral superconductivity on a silicon surface},\ }\href {https://doi.org/10.1038/s41567-022-01889-1} {\bibfield  {journal} {\bibinfo  {journal} {Nature Physics}\ }\textbf
  {\bibinfo {volume} {19}},\ \bibinfo {pages} {500} (\bibinfo {year} {2023})}\BibitemShut {NoStop}%
\bibitem [{\citenamefont {Fu}\ and\ \citenamefont {Kane}(2008)}]{Fu2008superconducting}%
  \BibitemOpen
  \bibfield  {author} {\bibinfo {author} {\bibfnamefont {L.}~\bibnamefont {Fu}}\ and\ \bibinfo {author} {\bibfnamefont {C.~L.}\ \bibnamefont {Kane}},\ }\bibfield  {title} {\bibinfo {title} {Superconducting proximity effect and majorana fermions at the surface of a topological insulator},\ }\href {https://doi.org/10.1103/PhysRevLett.100.096407} {\bibfield  {journal} {\bibinfo  {journal} {Phys. Rev. Lett.}\ }\textbf {\bibinfo {volume} {100}},\ \bibinfo {pages} {096407} (\bibinfo {year} {2008})}\BibitemShut {NoStop}%
\bibitem [{\citenamefont {Lutchyn}\ \emph {et~al.}(2010)\citenamefont {Lutchyn}, \citenamefont {Sau},\ and\ \citenamefont {Das~Sarma}}]{Lutchyn2010Majorana}%
  \BibitemOpen
  \bibfield  {author} {\bibinfo {author} {\bibfnamefont {R.~M.}\ \bibnamefont {Lutchyn}}, \bibinfo {author} {\bibfnamefont {J.~D.}\ \bibnamefont {Sau}},\ and\ \bibinfo {author} {\bibfnamefont {S.}~\bibnamefont {Das~Sarma}},\ }\bibfield  {title} {\bibinfo {title} {Majorana fermions and a topological phase transition in semiconductor-superconductor heterostructures},\ }\href {https://doi.org/10.1103/PhysRevLett.105.077001} {\bibfield  {journal} {\bibinfo  {journal} {Phys. Rev. Lett.}\ }\textbf {\bibinfo {volume} {105}},\ \bibinfo {pages} {077001} (\bibinfo {year} {2010})}\BibitemShut {NoStop}%
\bibitem [{\citenamefont {Yi}\ \emph {et~al.}(2024)\citenamefont {Yi}, \citenamefont {Zhao}, \citenamefont {Chan}, \citenamefont {Cai}, \citenamefont {Mei}, \citenamefont {Wu}, \citenamefont {Yan}, \citenamefont {Zhou}, \citenamefont {Zhang}, \citenamefont {Wang}, \citenamefont {Paolini}, \citenamefont {Xiao}, \citenamefont {Wang}, \citenamefont {Richardella}, \citenamefont {Singleton}, \citenamefont {Winter}, \citenamefont {Prokscha}, \citenamefont {Salman}, \citenamefont {Suter}, \citenamefont {Balakrishnan}, \citenamefont {Grutter}, \citenamefont {Chan}, \citenamefont {Samarth}, \citenamefont {Xu}, \citenamefont {Wu}, \citenamefont {Liu},\ and\ \citenamefont {Chang}}]{Hemian2024interface}%
  \BibitemOpen
  \bibfield  {author} {\bibinfo {author} {\bibfnamefont {H.}~\bibnamefont {Yi}}, \bibinfo {author} {\bibfnamefont {Y.-F.}\ \bibnamefont {Zhao}}, \bibinfo {author} {\bibfnamefont {Y.-T.}\ \bibnamefont {Chan}}, \bibinfo {author} {\bibfnamefont {J.}~\bibnamefont {Cai}}, \bibinfo {author} {\bibfnamefont {R.}~\bibnamefont {Mei}}, \bibinfo {author} {\bibfnamefont {X.}~\bibnamefont {Wu}}, \bibinfo {author} {\bibfnamefont {Z.-J.}\ \bibnamefont {Yan}}, \bibinfo {author} {\bibfnamefont {L.-J.}\ \bibnamefont {Zhou}}, \bibinfo {author} {\bibfnamefont {R.}~\bibnamefont {Zhang}}, \bibinfo {author} {\bibfnamefont {Z.}~\bibnamefont {Wang}}, \bibinfo {author} {\bibfnamefont {S.}~\bibnamefont {Paolini}}, \bibinfo {author} {\bibfnamefont {R.}~\bibnamefont {Xiao}}, \bibinfo {author} {\bibfnamefont {K.}~\bibnamefont {Wang}}, \bibinfo {author} {\bibfnamefont {A.~R.}\ \bibnamefont {Richardella}}, \bibinfo {author} {\bibfnamefont {J.}~\bibnamefont {Singleton}}, \bibinfo {author} {\bibfnamefont {L.~E.}\ \bibnamefont {Winter}},
  \bibinfo {author} {\bibfnamefont {T.}~\bibnamefont {Prokscha}}, \bibinfo {author} {\bibfnamefont {Z.}~\bibnamefont {Salman}}, \bibinfo {author} {\bibfnamefont {A.}~\bibnamefont {Suter}}, \bibinfo {author} {\bibfnamefont {P.~P.}\ \bibnamefont {Balakrishnan}}, \bibinfo {author} {\bibfnamefont {A.~J.}\ \bibnamefont {Grutter}}, \bibinfo {author} {\bibfnamefont {M.~H.~W.}\ \bibnamefont {Chan}}, \bibinfo {author} {\bibfnamefont {N.}~\bibnamefont {Samarth}}, \bibinfo {author} {\bibfnamefont {X.}~\bibnamefont {Xu}}, \bibinfo {author} {\bibfnamefont {W.}~\bibnamefont {Wu}}, \bibinfo {author} {\bibfnamefont {C.-X.}\ \bibnamefont {Liu}},\ and\ \bibinfo {author} {\bibfnamefont {C.-Z.}\ \bibnamefont {Chang}},\ }\bibfield  {title} {\bibinfo {title} {Interface-induced superconductivity in magnetic topological insulators},\ }\href {https://doi.org/10.1126/science.adk1270} {\bibfield  {journal} {\bibinfo  {journal} {Science}\ }\textbf {\bibinfo {volume} {383}},\ \bibinfo {pages} {634} (\bibinfo {year} {2024})}\BibitemShut
  {NoStop}%
\bibitem [{\citenamefont {Kivelson}(2002)}]{Kivelson2002}%
  \BibitemOpen
  \bibfield  {author} {\bibinfo {author} {\bibfnamefont {S.}~\bibnamefont {Kivelson}},\ }\bibfield  {title} {\bibinfo {title} {Making high {$T_c$} higher: a theoretical proposal},\ }\href {https://doi.org/https://doi.org/10.1016/S0921-4526(02)00775-5} {\bibfield  {journal} {\bibinfo  {journal} {Physica B: Condensed Matter}\ }\textbf {\bibinfo {volume} {318}},\ \bibinfo {pages} {61} (\bibinfo {year} {2002})},\ \bibinfo {note} {the Future of Materials Physics: A Festschrift for Zachary Fisk}\BibitemShut {NoStop}%
\bibitem [{\citenamefont {Berg}\ \emph {et~al.}(2008)\citenamefont {Berg}, \citenamefont {Orgad},\ and\ \citenamefont {Kivelson}}]{berg2008route}%
  \BibitemOpen
  \bibfield  {author} {\bibinfo {author} {\bibfnamefont {E.}~\bibnamefont {Berg}}, \bibinfo {author} {\bibfnamefont {D.}~\bibnamefont {Orgad}},\ and\ \bibinfo {author} {\bibfnamefont {S.~A.}\ \bibnamefont {Kivelson}},\ }\bibfield  {title} {\bibinfo {title} {Route to high-temperature superconductivity in composite systems},\ }\href {https://doi.org/10.1103/PhysRevB.78.094509} {\bibfield  {journal} {\bibinfo  {journal} {Phys. Rev. B}\ }\textbf {\bibinfo {volume} {78}},\ \bibinfo {pages} {094509} (\bibinfo {year} {2008})}\BibitemShut {NoStop}%
\bibitem [{\citenamefont {Wachtel}\ \emph {et~al.}(2012)\citenamefont {Wachtel}, \citenamefont {Bar-Yaacov},\ and\ \citenamefont {Orgad}}]{wachtel2012superfluid}%
  \BibitemOpen
  \bibfield  {author} {\bibinfo {author} {\bibfnamefont {G.}~\bibnamefont {Wachtel}}, \bibinfo {author} {\bibfnamefont {A.}~\bibnamefont {Bar-Yaacov}},\ and\ \bibinfo {author} {\bibfnamefont {D.}~\bibnamefont {Orgad}},\ }\bibfield  {title} {\bibinfo {title} {Superfluid stiffness renormalization and critical temperature enhancement in a composite superconductor},\ }\href {https://doi.org/10.1103/PhysRevB.86.134531} {\bibfield  {journal} {\bibinfo  {journal} {Phys. Rev. B}\ }\textbf {\bibinfo {volume} {86}},\ \bibinfo {pages} {134531} (\bibinfo {year} {2012})}\BibitemShut {NoStop}%
\bibitem [{\citenamefont {Zujev}\ \emph {et~al.}(2014)\citenamefont {Zujev}, \citenamefont {Scalettar}, \citenamefont {Batrouni},\ and\ \citenamefont {Sengupta}}]{zujev2014pairing}%
  \BibitemOpen
  \bibfield  {author} {\bibinfo {author} {\bibfnamefont {A.}~\bibnamefont {Zujev}}, \bibinfo {author} {\bibfnamefont {R.~T.}\ \bibnamefont {Scalettar}}, \bibinfo {author} {\bibfnamefont {G.~G.}\ \bibnamefont {Batrouni}},\ and\ \bibinfo {author} {\bibfnamefont {P.}~\bibnamefont {Sengupta}},\ }\bibfield  {title} {\bibinfo {title} {Pairing correlations in the two-layer attractive {H}ubbard model},\ }\href {https://iopscience.iop.org/article/10.1088/1367-2630/16/1/013004/pdf} {\bibfield  {journal} {\bibinfo  {journal} {New Journal of Physics}\ }\textbf {\bibinfo {volume} {16}},\ \bibinfo {pages} {013004} (\bibinfo {year} {2014})}\BibitemShut {NoStop}%
\bibitem [{\citenamefont {Dee}\ \emph {et~al.}(2022)\citenamefont {Dee}, \citenamefont {Johnston},\ and\ \citenamefont {Maier}}]{dee2022enhancing}%
  \BibitemOpen
  \bibfield  {author} {\bibinfo {author} {\bibfnamefont {P.~M.}\ \bibnamefont {Dee}}, \bibinfo {author} {\bibfnamefont {S.}~\bibnamefont {Johnston}},\ and\ \bibinfo {author} {\bibfnamefont {T.~A.}\ \bibnamefont {Maier}},\ }\bibfield  {title} {\bibinfo {title} {Enhancing {${T}_{\mathrm{c}}$} in a composite superconductor/metal bilayer system: A dynamical cluster approximation study},\ }\href {https://doi.org/10.1103/PhysRevB.105.214502} {\bibfield  {journal} {\bibinfo  {journal} {Phys. Rev. B}\ }\textbf {\bibinfo {volume} {105}},\ \bibinfo {pages} {214502} (\bibinfo {year} {2022})}\BibitemShut {NoStop}%
\bibitem [{\citenamefont {Fontenele}\ \emph {et~al.}(2024)\citenamefont {Fontenele}, \citenamefont {Costa}, \citenamefont {Paiva},\ and\ \citenamefont {dos Santos}}]{bilayerattrhub}%
  \BibitemOpen
  \bibfield  {author} {\bibinfo {author} {\bibfnamefont {R.~A.}\ \bibnamefont {Fontenele}}, \bibinfo {author} {\bibfnamefont {N.~C.}\ \bibnamefont {Costa}}, \bibinfo {author} {\bibfnamefont {T.}~\bibnamefont {Paiva}},\ and\ \bibinfo {author} {\bibfnamefont {R.~R.}\ \bibnamefont {dos Santos}},\ }\bibfield  {title} {\bibinfo {title} {Increasing superconducting ${T}_{c}$ by layering in the attractive hubbard model},\ }\href {https://doi.org/10.1103/PhysRevA.110.053315} {\bibfield  {journal} {\bibinfo  {journal} {Phys. Rev. A}\ }\textbf {\bibinfo {volume} {110}},\ \bibinfo {pages} {053315} (\bibinfo {year} {2024})}\BibitemShut {NoStop}%
\bibitem [{\citenamefont {Hettler}\ \emph {et~al.}(2000)\citenamefont {Hettler}, \citenamefont {Mukherjee}, \citenamefont {Jarrell},\ and\ \citenamefont {Krishnamurthy}}]{hettler2000dynamical}%
  \BibitemOpen
  \bibfield  {author} {\bibinfo {author} {\bibfnamefont {M.}~\bibnamefont {Hettler}}, \bibinfo {author} {\bibfnamefont {M.}~\bibnamefont {Mukherjee}}, \bibinfo {author} {\bibfnamefont {M.}~\bibnamefont {Jarrell}},\ and\ \bibinfo {author} {\bibfnamefont {H.}~\bibnamefont {Krishnamurthy}},\ }\bibfield  {title} {\bibinfo {title} {Dynamical cluster approximation: Nonlocal dynamics of correlated electron systems},\ }\href {https://journals.aps.org/prb/pdf/10.1103/PhysRevB.61.12739} {\bibfield  {journal} {\bibinfo  {journal} {Physical Review B}\ }\textbf {\bibinfo {volume} {61}},\ \bibinfo {pages} {12739} (\bibinfo {year} {2000})}\BibitemShut {NoStop}%
\bibitem [{\citenamefont {Maier}\ \emph {et~al.}(2005)\citenamefont {Maier}, \citenamefont {Jarrell}, \citenamefont {Pruschke},\ and\ \citenamefont {Hettler}}]{maier2005quantum}%
  \BibitemOpen
  \bibfield  {author} {\bibinfo {author} {\bibfnamefont {T.}~\bibnamefont {Maier}}, \bibinfo {author} {\bibfnamefont {M.}~\bibnamefont {Jarrell}}, \bibinfo {author} {\bibfnamefont {T.}~\bibnamefont {Pruschke}},\ and\ \bibinfo {author} {\bibfnamefont {M.~H.}\ \bibnamefont {Hettler}},\ }\bibfield  {title} {\bibinfo {title} {Quantum cluster theories},\ }\href {https://journals.aps.org/rmp/pdf/10.1103/RevModPhys.77.1027} {\bibfield  {journal} {\bibinfo  {journal} {Reviews of Modern Physics}\ }\textbf {\bibinfo {volume} {77}},\ \bibinfo {pages} {1027} (\bibinfo {year} {2005})}\BibitemShut {NoStop}%
\bibitem [{\citenamefont {Cohen-Stead}\ \emph {et~al.}(2024{\natexlab{a}})\citenamefont {Cohen-Stead}, \citenamefont {{Malkaruge Costa}}, \citenamefont {Neuhaus}, \citenamefont {{Tanjaroon Ly}}, \citenamefont {Zhang}, \citenamefont {Scalettar}, \citenamefont {Barros},\ and\ \citenamefont {Johnston}}]{smoqy}%
  \BibitemOpen
  \bibfield  {author} {\bibinfo {author} {\bibfnamefont {B.}~\bibnamefont {Cohen-Stead}}, \bibinfo {author} {\bibfnamefont {S.}~\bibnamefont {{Malkaruge Costa}}}, \bibinfo {author} {\bibfnamefont {J.}~\bibnamefont {Neuhaus}}, \bibinfo {author} {\bibfnamefont {A.}~\bibnamefont {{Tanjaroon Ly}}}, \bibinfo {author} {\bibfnamefont {Y.}~\bibnamefont {Zhang}}, \bibinfo {author} {\bibfnamefont {R.}~\bibnamefont {Scalettar}}, \bibinfo {author} {\bibfnamefont {K.}~\bibnamefont {Barros}},\ and\ \bibinfo {author} {\bibfnamefont {S.}~\bibnamefont {Johnston}},\ }\bibfield  {title} {\bibinfo {title} {{SmoQyDQMC.jl: A flexible implementation of determinant quantum Monte Carlo for Hubbard and electron-phonon interactions}},\ }\href {https://doi.org/10.21468/SciPostPhysCodeb.29} {\bibfield  {journal} {\bibinfo  {journal} {SciPost Phys. Codebases}\ ,\ \bibinfo {pages} {29}} (\bibinfo {year} {2024}{\natexlab{a}})}\BibitemShut {NoStop}%
\bibitem [{\citenamefont {Cohen-Stead}\ \emph {et~al.}(2024{\natexlab{b}})\citenamefont {Cohen-Stead}, \citenamefont {{Malkaruge Costa}}, \citenamefont {Neuhaus}, \citenamefont {{Tanjaroon Ly}}, \citenamefont {Zhang}, \citenamefont {Scalettar}, \citenamefont {Barros},\ and\ \citenamefont {Johnston}}]{smoqy_code}%
  \BibitemOpen
  \bibfield  {author} {\bibinfo {author} {\bibfnamefont {B.}~\bibnamefont {Cohen-Stead}}, \bibinfo {author} {\bibfnamefont {S.}~\bibnamefont {{Malkaruge Costa}}}, \bibinfo {author} {\bibfnamefont {J.}~\bibnamefont {Neuhaus}}, \bibinfo {author} {\bibfnamefont {A.}~\bibnamefont {{Tanjaroon Ly}}}, \bibinfo {author} {\bibfnamefont {Y.}~\bibnamefont {Zhang}}, \bibinfo {author} {\bibfnamefont {R.}~\bibnamefont {Scalettar}}, \bibinfo {author} {\bibfnamefont {K.}~\bibnamefont {Barros}},\ and\ \bibinfo {author} {\bibfnamefont {S.}~\bibnamefont {Johnston}},\ }\bibfield  {title} {\bibinfo {title} {{Codebase release r0.3 for SmoQyDQMC.jl}},\ }\href {https://doi.org/10.21468/SciPostPhysCodeb.29-r0.3} {\bibfield  {journal} {\bibinfo  {journal} {SciPost Phys. Codebases}\ ,\ \bibinfo {pages} {29}} (\bibinfo {year} {2024}{\natexlab{b}})}\BibitemShut {NoStop}%
\bibitem [{\citenamefont {Scalapino}\ \emph {et~al.}(1993)\citenamefont {Scalapino}, \citenamefont {White},\ and\ \citenamefont {Zhang}}]{SWZ}%
  \BibitemOpen
  \bibfield  {author} {\bibinfo {author} {\bibfnamefont {D.~J.}\ \bibnamefont {Scalapino}}, \bibinfo {author} {\bibfnamefont {S.~R.}\ \bibnamefont {White}},\ and\ \bibinfo {author} {\bibfnamefont {S.}~\bibnamefont {Zhang}},\ }\bibfield  {title} {\bibinfo {title} {Insulator, metal, or superconductor: The criteria},\ }\href {https://journals.aps.org/prb/abstract/10.1103/PhysRevB.47.7995} {\bibfield  {journal} {\bibinfo  {journal} {Physical Review B}\ }\textbf {\bibinfo {volume} {47}},\ \bibinfo {pages} {7995} (\bibinfo {year} {1993})}\BibitemShut {NoStop}%
\bibitem [{\citenamefont {Fontenele}\ \emph {et~al.}(2022)\citenamefont {Fontenele}, \citenamefont {Costa}, \citenamefont {dos Santos},\ and\ \citenamefont {Paiva}}]{attrHub_BCS-BEC}%
  \BibitemOpen
  \bibfield  {author} {\bibinfo {author} {\bibfnamefont {R.~A.}\ \bibnamefont {Fontenele}}, \bibinfo {author} {\bibfnamefont {N.~C.}\ \bibnamefont {Costa}}, \bibinfo {author} {\bibfnamefont {R.~R.}\ \bibnamefont {dos Santos}},\ and\ \bibinfo {author} {\bibfnamefont {T.}~\bibnamefont {Paiva}},\ }\bibfield  {title} {\bibinfo {title} {Two-dimensional attractive {H}ubbard model and the {BCS}-{BEC} crossover},\ }\href {https://journals.aps.org/prb/abstract/10.1103/PhysRevB.105.184502} {\bibfield  {journal} {\bibinfo  {journal} {Phys. Rev. B}\ }\textbf {\bibinfo {volume} {105}},\ \bibinfo {pages} {184502} (\bibinfo {year} {2022})}\BibitemShut {NoStop}%
\bibitem [{\citenamefont {Scalettar}\ \emph {et~al.}(1989)\citenamefont {Scalettar}, \citenamefont {Loh}, \citenamefont {Gubernatis}, \citenamefont {Moreo}, \citenamefont {White}, \citenamefont {Scalapino}, \citenamefont {Sugar},\ and\ \citenamefont {Dagotto}}]{scalettar1989phase}%
  \BibitemOpen
  \bibfield  {author} {\bibinfo {author} {\bibfnamefont {R.~T.}\ \bibnamefont {Scalettar}}, \bibinfo {author} {\bibfnamefont {E.~Y.}\ \bibnamefont {Loh}}, \bibinfo {author} {\bibfnamefont {J.~E.}\ \bibnamefont {Gubernatis}}, \bibinfo {author} {\bibfnamefont {A.}~\bibnamefont {Moreo}}, \bibinfo {author} {\bibfnamefont {S.~R.}\ \bibnamefont {White}}, \bibinfo {author} {\bibfnamefont {D.~J.}\ \bibnamefont {Scalapino}}, \bibinfo {author} {\bibfnamefont {R.~L.}\ \bibnamefont {Sugar}},\ and\ \bibinfo {author} {\bibfnamefont {E.}~\bibnamefont {Dagotto}},\ }\bibfield  {title} {\bibinfo {title} {Phase diagram of the two-dimensional negative-{$U$} {H}ubbard model},\ }\href {https://doi.org/10.1103/PhysRevLett.62.1407} {\bibfield  {journal} {\bibinfo  {journal} {Phys. Rev. Lett.}\ }\textbf {\bibinfo {volume} {62}},\ \bibinfo {pages} {1407} (\bibinfo {year} {1989})}\BibitemShut {NoStop}%
\end{thebibliography}%

\end{document}